\documentclass[journal]{IEEEtran}

\usepackage{amsmath,amssymb,amsfonts,amsthm}
\usepackage{bm}
\usepackage{cite}
\usepackage{xcolor}
\usepackage{algorithm}
\usepackage{algpseudocode}
\usepackage{graphicx}
\usepackage{subcaption}
\usepackage[font=footnotesize]{caption}
\newtheorem{theorem}{Theorem}

\newcommand{\Nset}{\mathcal{N}}
\newcommand{\Kset}{\mathcal{K}}
\newcommand{\Rset}{\mathcal{R}}

\title{Joint Air Corridor and BS Coverage Planning for Large-Scale High-Density Low-Altitude Aerial Networks via Channel Knowledge Maps}

\author{Jiaxuan~Li, Yilong~Chen, Yuan~Guo, Fan~Liu,~\IEEEmembership{Senior Member,~IEEE}, and Jie~Xu,~\IEEEmembership{Fellow,~IEEE}%
\thanks{J. Li, Y. Chen, Y. Guo, and J. Xu are with the School of
Science and Engineering (SSE), Shenzhen Future Network of Intelligence
Institute (FNii-Shenzhen), and Guangdong Provincial Key Laboratory of
Future Networks of Intelligence, The Chinese University of Hong Kong,
Shenzhen, Guangdong 518172, China (e-mail: jiaxuanli4@link.cuhk.edu.cn; \mbox{chenyilong@cuhk.edu.cn}; guoyuan@cuhk.edu.cn; xujie@cuhk.edu.cn). 

F. Liu is with the National Mobile Communications Research Laboratory,
School of Information Science and Engineering, Southeast University, Nanjing
210096, China (e-mail: f.liu@ieee.org). 

F. Liu is the corresponding author.}}%Corresponding author: \TBD{confirm corresponding author and e-mail}.}}
\begin{document}
\bstctlcite{IEEEtran:BSTcontrol}

\maketitle

\begin{abstract}
This paper studies the joint air corridor and base station (BS) coverage planning problem for high-density low-altitude aerial networks supported by terrestrial cellular infrastructures. In the considered system, unmanned aerial vehicles (UAVs) operate within a structured air corridor planned over a grid-based airspace, and each BS serves selected corridor grid cells by sharing its communication resources among all UAVs therein. Each BS’s traffic load depends on both the number of covered corridor cells and the UAV density, thereby affecting the communication rate available to each UAV. We exploit the channel knowledge map (CKM) to characterize the channel gains between BSs and candidate airspace cells and evaluate desired signals and inter-cell interference for environment-aware offline planning. Based on CKM, we jointly optimize the corridor layout and BS coverage to minimize a weighted objective balancing corridor length and the supportable average UAV density along the entire corridor, subject to minimum communication rate requirements for all UAVs along the air corridor. The joint design is formulated as a mixed-integer linear program (MILP), which can be optimally solved using off-the-shelf solvers. For large maps, we develop a scalable suboptimal algorithm based on Lagrangian relaxation and A* search. Specifically, we incorporate BS resource constraints into A* corridor search via Lagrangian relaxation, and then jointly optimize BS coverage and UAV density along the generated corridors under the original rate constraints. Numerical results demonstrate the benefits of the proposed joint design in supporting high-density UAV operations and show that the proposed algorithm obtains high-quality solutions with substantially reduced computation time.
\end{abstract}

\begin{IEEEkeywords}
Channel knowledge map, air corridor planning, BS coverage, high-density traffic, mixed-integer programming.
\end{IEEEkeywords}

\section{Introduction}
\label{sec:introduction}

Urban air mobility, logistics delivery, emergency response, and infrastructure inspection are driving a growing demand for low-altitude aerial services \cite{wang2025laenetworks,xu2025emergency}. As these services expand, large numbers of unmanned aerial vehicles (UAVs) are expected to operate simultaneously within limited urban airspace, motivating the development of high-density low-altitude aerial networks. In such networks, UAV movements need to be coordinated to maintain safe separation and manage potential trajectory conflicts. This requires timely exchange of UAV positions, flight status, and intended trajectories, together with reliable delivery of control commands \cite{kopardekar2016utm}. Structured air corridors provide an effective means of organizing these operations by confining UAV movements to predefined three-dimensional (3D) flight routes. With specified flight directions and separation rules, these corridors facilitate the traffic flow coordination and the orderly use of limited airspace \cite{muna2021corridors}. Supporting dense UAV traffic within these corridors also places substantial demands on the wireless communication network. Wireless links need to support both control signaling and mission-related data transmission~\cite{zeng2019accessing}, while the available radio resources must be shared among multiple UAVs. As UAV density increases, the aggregate communication demand within a given airspace grows, making reliable connectivity and sufficient communication capacity essential for sustained large-scale UAV operations \cite{geraci2022future,ge2023systematic}.

Terrestrial cellular networks are a promising platform for providing wireless connectivity along air corridors, owing to their widely deployed base stations (BSs), licensed spectrum, and established mobility management mechanisms \cite{zeng2019cellular}. By accessing the cellular network as aerial users, UAVs can exchange control messages and mission data with ground entities while moving across the coverage areas of different BSs. Air corridors organize UAV traffic within structured regions of the airspace, facilitating targeted cellular coverage planning \cite{benzaghta2025corridors}. However, the channel gains between terrestrial BSs and aerial locations vary with propagation distance and environmental effects such as blockage and reflection. The resulting spatial variations in desired signal and inter-cell interference powers lead to location-dependent communication quality along potential corridors \cite{geraci2022future,bernabe2024massive}. For dense corridor operations, each BS must further share its limited communication resources among the UAVs it serves. Consequently, the achievable rate of each UAV depends on both its channel conditions and the traffic load of its serving BS. Different corridor layouts and BS coverage assignments can therefore lead to different resource demands across BSs and different levels of supportable UAV traffic.

Existing studies have investigated network optimization and UAV mobility design to improve cellular support for aerial operations. For given UAV corridors, BS antenna configuration, association, and network deployment have been studied to enhance aerial coverage and capacity \cite{maeng2024strongest,karimi2025capacity,karimi2024quantization}. Bayesian optimization has also been applied to BS antenna settings to improve corridor connectivity while accounting for terrestrial services \cite{benzaghta2025corridors}. For UAV missions, trajectory optimization has been used to maximize the minimum signal-to-interference-plus-noise ratio (SINR) or reduce handovers under communication quality requirements \cite{qin2024coverage,du2025handover}. Deep reinforcement learning has further been employed for joint speed control and BS association of multiple UAVs operating on aerial highways \cite{yan2023multiUAV}. Furthermore, networked integrated sensing and communication (ISAC) systems can be used to support UAVs, and the UAV trajectories, BS association, and coordinated transmit beamforming can be jointly optimized under sensing and communication requirements \cite{cheng2025networked}. These studies mainly focus on improving communication performance along given corridors or optimizing UAV operations for specific missions. However, joint planning of air corridors and BS coverage for supporting dense UAV traffic remains less explored.

To efficiently design the network, the channel information between BSs and candidate aerial locations is needed to evaluate the communication service
capabilities of different air corridor and BS coverage designs. Channel knowledge map (CKM) has emerged as a promising technique to acquire such information. CKM is a database for a given
geographical region that provides channel knowledge indexed
by transceiver locations. Such maps
can be obtained offline from environmental information and
limited channel measurements through ray tracing,
interpolation, or learning methods
\cite{zeng2024tutorial, ren2026ckmconstruction, wu2025ckmimagenet}. Compared with conventional
statistical channel models, CKM can exploit prior knowledge
of the local propagation environment to support
environment-aware network planning. CKM may include various
types of maps according to the channel knowledge they provide,
among which channel gain maps (CGMs) characterize large-scale
fading at different locations and provide channel power gains
for offline coverage planning. Existing studies have exploited
spatial channel knowledge for UAV placement
\cite{li2023multiUAV}, trajectory and transmission optimization
\cite{zhang2021radiomap,zhan2024video}, and beamforming and
user association along given air corridors
\cite{tarafder2025digitaltwin}. These studies mainly focus on
communication and operational performance for given UAVs
or flight missions. However, joint planning of air corridors
and BS coverage using CKM to increase the UAV density
supportable by existing cellular infrastructure remains
less explored.

In prior work~\cite{li2026ckm}, air corridors and BS
deployment were jointly optimized to minimize the system
cost comprising corridor length and the number of deployed
BSs, subject to communication and sensing coverage
constraints. The present work considers an existing cellular
network and treats the supportable UAV traffic density as
an optimization variable. In particular, we replace the
communication SINR constraints with minimum communication
rate constraints for each UAV under dense aerial traffic.
These constraints explicitly account for resource sharing
among all UAVs served by the same BS and therefore depend
jointly on corridor layout, BS coverage, and traffic density
through the aggregate BS load.

Motivated by the above, we exploit CGMs to obtain prior channel power gains between terrestrial BSs
and candidate airspace locations for evaluating desired
signal and interference powers. Next, we jointly optimize
the corridor layout, BS coverage, and aerial traffic density to minimize a weighted objective balancing corridor length and supportable traffic density, while
ensuring that each UAV along the corridor achieves its
required minimum communication rate.
Our main contributions are summarized as follows.

\begin{itemize}

    \item We propose a joint air corridor and BS coverage design
    using CKM to optimize the corridor layout, the grid cells
    covered by each BS, and the normalized aerial traffic
    density. The design accounts for spatial channel conditions
    and aggregate BS loads to balance corridor length and
    supportable traffic density while meeting the minimum rate
    requirement of each UAV.

    \item We establish an exact mixed-integer linear programming
    (MILP) formulation of the joint
    planning problem. By exploiting the coverage of each
    selected grid cell by a single BS, we equivalently
    reformulate the rate constraints and exactly linearize
    the products between traffic density and binary coverage
    variables, enabling globally optimal solutions for
    small maps. We further prove that the adopted
    connectivity constraints, together with a positive
    corridor length term in the objective, exclude isolated
    cycles from globally optimal solutions.

    \item We develop a scalable algorithm based on Lagrangian
    relaxation and A* search for large maps. For each trial
    traffic density, we relax the original rate constraints
    into additive BS resource constraints and use Lagrangian
    multipliers to incorporate resource consumption into
    the A* search costs, generating candidate corridors.
    For each given corridor, we then jointly optimize BS
    coverage and continuous traffic density under the
    original equal resource sharing model to recover a
    feasible solution satisfying the rate requirements.
    We further analyze the convergence of the best dual
    value and the computational complexity of the algorithm.
    
    \item Numerical results demonstrate the benefit of jointly planning corridor layouts and BS coverage for increasing the supportable UAV density. The proposed algorithm achieves lower objective values than the considered benchmarks and comparable traffic densities to the full MILP on the tested small maps.

\end{itemize}

The remainder of this paper is organized as follows. Section II introduces the low-altitude airspace model, the CKM-assisted channel representation, and the cellular communication and aerial traffic-load models. Section III formulates the joint air corridor planning, BS coverage, and aerial traffic-density optimization problem. Section IV presents the exact MILP solution for small-scale maps. Section V develops the scalable Lagrangian--A* algorithm for large-scale maps and analyzes its theoretical properties and computational complexity. Section VI provides numerical results to evaluate the proposed design. Section VII concludes this paper.

\emph{Notations.}
Vectors are denoted by boldface lowercase letters.
The superscript $(\cdot)^{\mathsf T}$ denotes transpose,
and $\|\cdot\|_2$ denotes the Euclidean norm.
The set of real numbers is denoted by $\mathbb R$.
For a finite set $\mathcal S$, $|\mathcal S|$ denotes
its cardinality. For a real number $a$, its positive
part is defined as $[a]^+\triangleq\max\{a,0\}$.

\section{System Model}
\label{sec:model}

\begin{figure}[t]
    \centering
    \includegraphics[width=\columnwidth]{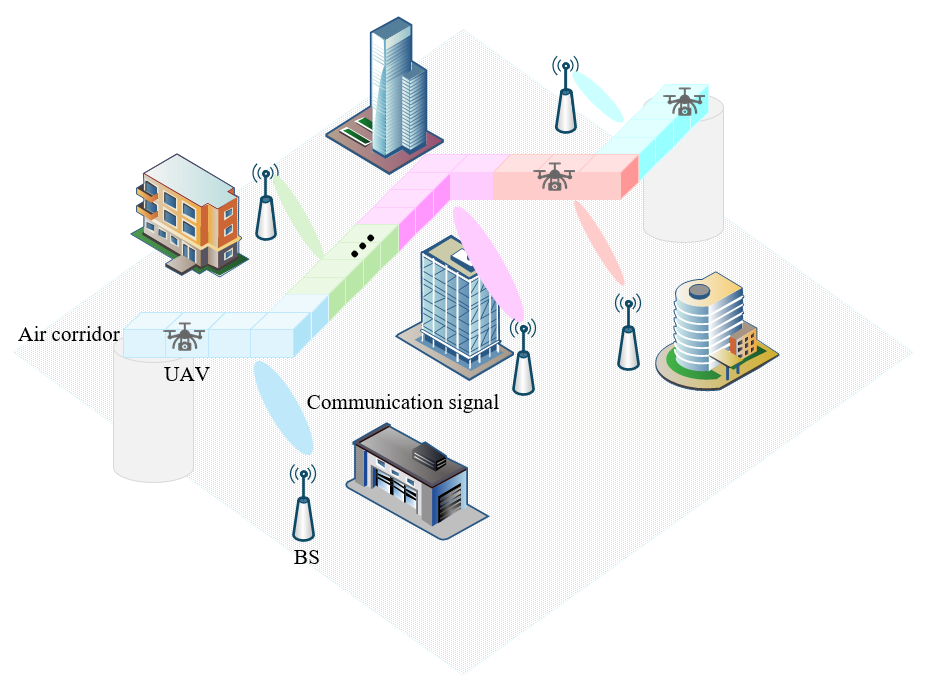}
    \caption{Illustration of the considered low-altitude aerial network.}
    \label{fig:system_model}
\end{figure}

This section presents the considered low-altitude aerial network in which UAVs operate within a structured air corridor and are served by
existing terrestrial BSs, as illustrated in Fig.~\ref{fig:system_model}. The considered airspace is represented by 3D grid cells, and the corridor consists of connected cells between prescribed departure and destination cells. Each corridor cell is covered by one BS, whose communication resources are shared among all UAVs it serves. We jointly plan the corridor layout and BS coverage to support dense aerial traffic while satisfying the minimum communication rate requirement of each UAV.\footnote{UAV positioning and collision avoidance are assumed to be
supported by onboard navigation and sensing systems, such
as global navigation satellite system (GNSS) receivers,
inertial sensors, and cameras, while
traffic coordination is handled by a dedicated traffic
management system. This work focuses on the communication
coverage and capacity provided by BSs along
the planned air corridor.}

\subsection{Low-Altitude Airspace and BS Network}
We consider UAV operations within a 3D airspace
layer over an urban area. The layer is centered at altitude $H$
and has a prescribed vertical extent $\Delta_z$. Following the
grid representation adopted in \cite{li2026ckm}, the airspace
is discretized into an $N\times N$ arrangement of cuboidal grid
cells, with dimensions $\Delta_x$, $\Delta_y$, and $\Delta_z$
along the $x$, $y$, and $z$ directions, respectively.
We define $\Nset\triangleq\{1,\ldots,N\}$ and consider equal
horizontal dimensions $\Delta_x=\Delta_y\triangleq\Delta$.

Let $(x_{\rm O},y_{\rm O})$ denote the horizontal reference
coordinate of the considered region. The $(i,j)$-th airspace
grid cell is defined as
\begin{equation}
\begin{aligned}
    \Rset_{i,j}
    ={}&
    [x_{\rm O}+(i-1)\Delta_x,\,
     x_{\rm O}+i\Delta_x]\\
    &\times
    [y_{\rm O}+(j-1)\Delta_y,\,
     y_{\rm O}+j\Delta_y]\\
    &\times
    [H-\Delta_z/2,\,
     H+\Delta_z/2],
\end{aligned}
\label{eq:grid_region}
\end{equation}
for $(i,j)\in\Nset^2$. Each selected grid cell represents a
3D corridor segment. The corridor layout is
determined by connecting these segments in the horizontal
directions, while their vertical extent is specified by
$\Delta_z$.

Let $s=(i_{\rm s},j_{\rm s})\in\Nset^2,$ $d=(i_{\rm d},j_{\rm d})\in\Nset^2$ denote the prescribed departure and destination cells, respectively. Since UAV movement is confined to the considered flight layer, two grid cells are regarded as adjacent if they share a common vertical face. Accordingly, for each $(i,j)\in\Nset^2$, its four-neighbor set is defined as
\begin{equation}
    \Nset_4(i,j)
    \triangleq
    \left\{
    (i',j')\in\Nset^2:
    |i'-i|+|j'-j|=1
    \right\}.
    \label{eq:neighbor_set}
\end{equation}
An air corridor connecting $s$ and $d$ is represented by a sequence of face-adjacent grid cells according to \eqref{eq:neighbor_set}.\footnote{
This work considers a single air corridor.
For multiple corridors at different altitudes, CKMs can
characterize the channels in the corresponding airspace
layers, while the load of each BS must include the UAVs
it serves across all corridors. The joint design of
multiple corridors and their BS coverage is left for
future work.} Since the grid resolution is uniform, each transition between two adjacent cells corresponds to a horizontal distance of $\Delta$, and hence the corridor length is proportional to the number of traversed grid cells.

The considered airspace is served by an existing terrestrial cellular network consisting of $K$ BSs indexed by $\Kset\triangleq\{1,\ldots,K\}$. Each BS is equipped with a single antenna. The 3D locations of all BSs are fixed and known.
Due to the site-specific propagation environment, different airspace grid cells may experience substantially different channel conditions with respect to the terrestrial BSs. Accordingly, the cellular service capability over each grid will be characterized based on the BS-dependent location-specific channel information introduced in the next subsection.

\subsection{CKM-Based Channel Representation}

To characterize the propagation conditions over the considered
airspace, we employ CGMs, a particular type
of CKM that provides channel power gains indexed by transceiver
locations \cite{zeng2024tutorial}. Consider a receiver region $\mathcal D\subset\mathbb R^3$ partitioned into 3D receiver grids according to the spatial resolution of the CGMs. Let $\mathcal Q$ denote the set of receiver grid centers, and let $\mathcal C(\bm q)\subset\mathcal D$ denote the receiver grid
centered at $\bm q\in\mathcal Q$. For BS $k$, let $h_k(\bm p)$ denote
the channel coefficient between BS $k$ and receiver location
$\bm p$. The CGM entry indexed by $\bm q$ is defined as
\begin{equation}
    \mathcal M_k(\bm q)
    \triangleq
    \frac{1}{|\mathcal C(\bm q)|}
    \int_{\mathcal C(\bm q)}
    |h_k(\bm p)|^2\,\mathrm dV(\bm p),
    \label{eq:cgm_receiver_gain}
\end{equation}
for $k\in\Kset$ and $\bm q\in\mathcal Q$, where
$|\mathcal C(\bm q)|$ denotes the receiver grid volume
and $\mathrm dV(\bm p)$ is the volume element.
Thus, each entry represents the average channel power gain
over a receiver grid and is indexed by its center
\cite{chen2026gscgm}.

Under the far field and local homogeneity assumptions,
the receiver grid dimensions are small relative to the
distances to dominant propagation structures, so that
path loss and propagation angles vary little within
each grid. The grid dimensions are also assumed to be
large compared with the carrier wavelength. Spatial
power averaging then reduces the influence of local
phase variations while preserving large-scale propagation
characteristics \cite{chen2026gscgm}. 

The CGMs can be obtained offline from channel measurements,
ray tracing, interpolation, or learning methods
\cite{zeng2024tutorial}. We assume that the CGMs are available
prior to corridor planning and treat their entries as
given inputs. For corridor planning, we retrieve CGM entries at
$N_{\rm q}$ receiver grid centers
$\bm q_{i,j}^{(n)}\in\mathcal Q\cap\Rset_{i,j}$,
$n=1,\ldots,N_{\rm q}$, within each planning cell.
These centers may have different horizontal positions
and altitudes. The corresponding locally averaged
channel power gain is denoted by
$g_{k,i,j}^{(n)}
\triangleq\mathcal M_k(\bm q_{i,j}^{(n)})$.
The same receiver grid centers are used for all BSs,
ensuring that desired and interfering channel gains
correspond to the same spatial regions.

\subsection{Cell-Level Communication Service Model}

Based on the CGM entries introduced above, we next characterize the
communication service capability of each BS over the airspace grid cells.
We consider a common, fixed bandwidth $B$ for aerial communication, reused across the BSs. Each BS transmits with a fixed power $P$.
UAVs served by the same BS are orthogonally scheduled in time or frequency, such that intra-cell interference is avoided. When evaluating the link between a UAV and its serving BS, signals from the other BSs over the same spectrum are treated as inter-cell interference.

Consider service by BS $k$ over the receiver grid centered at
$\bm q_{i,j}^{(n)}$. Using the locally averaged channel gains,
the corresponding equivalent received signal is
\begin{equation}
    y_{k,i,j}^{(n)}
    =
    \sqrt{P G g_{k,i,j}^{(n)}}\,\xi_k
    +
    \sum_{b\in\Kset\setminus\{k\}}
    \sqrt{P g_{b,i,j}^{(n)}}\,\xi_b
    + \nu_{k,i,j},
    \label{eq:received_signal}
\end{equation}
where $\xi_b$ denotes the unit-power effective symbol transmitted by
BS $b$, $G\ge1$ is a coefficient capturing the transmit antenna gain, and $\nu_{k,i,j}\sim\mathcal{CN}(0,N_0B)$ denotes the receiver noise. Accordingly, the effective SINR for service by BS $k$ at
$\bm q_{i,j}^{(n)}$ is
\begin{equation}
    \gamma_{k,i,j}^{(n)}
    =
    \frac{
        PG g_{k,i,j}^{(n)}
    }{
        \displaystyle
        \sum_{b\in\Kset\setminus\{k\}}
        P g_{b,i,j}^{(n)}
        + N_0 B
    }.
    \label{eq:pointwise_sinr}
\end{equation}

The desired and interfering channel gains in
\eqref{eq:pointwise_sinr} represent local averages over
the same receiver grid. To characterize propagation
variations within planning cell $\Rset_{i,j}$, we use
the minimum effective SINR among the CGM entries
retrieved within that cell. Specifically, we define
\begin{equation}
    \underline{\gamma}_{k,i,j}
    \triangleq
    \min_{n=1,\ldots,N_{\rm q}}
    \gamma_{k,i,j}^{(n)},
    \quad
    k\in\Kset,\;
    (i,j)\in\Nset^2,
    \label{eq:grid_sinr}
\end{equation}
as the effective SINR from BS $k$ to grid cell $(i,j)$. Accordingly, the spectral efficiency from BS $k$ to grid cell $(i,j)$ is defined as
\begin{equation}
    \eta_{k,i,j}
    \triangleq
    \log_2\!\left(
        1+\underline{\gamma}_{k,i,j}
    \right).
    \label{eq:grid_spectral_efficiency}
\end{equation}
Here, $\eta_{k,i,j}$ is measured in bit/s/Hz and characterizes the communication link between BS $k$ and grid cell $(i,j)$. Since it depends only on the offline CGMs and the fixed cellular network parameters, $\eta_{k,i,j}$ can be computed prior to corridor planning. For each UAV, its communication rate under resource sharing depends on the allocated bandwidth, as specified below.

\subsection{Aerial Traffic and Resource-Sharing Model}

We next characterize the aerial traffic demand along the air corridor and the corresponding communication rates provided
by the terrestrial BSs. Let $u$ denote the equivalent average
number of simultaneously active UAVs in each airspace grid
cell included in the corridor. Since all grid cells have
the same volume $\Delta_x\Delta_y\Delta_z$, $u$ is proportional
to the spatial density of active UAVs and is thus used as
the normalized aerial traffic density in the subsequent design.

We adopt a grid-level BS coverage model, where the UAVs located in the same corridor grid cell are served by a common BS. Let $x_{k,i,j}\in\{0,1\}$ denote the BS coverage indicator, where $x_{k,i,j}=1$ indicates that grid cell $(i,j)$ is covered by BS $k$, and $x_{k,i,j}=0$ otherwise. The coverage constraints for selected and unselected airspace grid cells will be specified in the subsequent problem formulation.

Since each selected grid cell contributes an average traffic load of $u$, the aggregate aerial-user load served by BS $k$ is
\begin{equation}
    u_k
    =
    u
    \sum_{i=1}^{N}\sum_{j=1}^{N}
    x_{k,i,j},
    \quad k\in\Kset.
    \label{eq:bs_user_load}
\end{equation}
Here, \(u_k\) aggregates the traffic loads of all corridor grid cells covered by BS \(k\) and is used to determine the resource sharing among the UAVs served by that BS.

With orthogonal scheduling and long-term equal resource sharing at each BS, the load-based rate estimate for a UAV in cell $(i,j)$ served by BS $k$ is
\begin{equation}
    r_{k,i,j}
    =
    \frac{B}{u_k}\eta_{k,i,j}.
    \label{eq:per_uav_rate}
\end{equation}

% \begin{table}[t]
% \centering
% \caption{Main Notations}
% \label{tab:main_notations}
% \footnotesize
% \renewcommand{\arraystretch}{1.08}
% \setlength{\tabcolsep}{4pt}
% \begin{tabular}{p{0.27\columnwidth} p{0.64\columnwidth}}
% \hline
% \textbf{Notation} & \textbf{Definition} \\
% \hline
% $\Nset$ & Grid index set \\
% $N$ & Number of grids per dimension \\
% $\Delta_x,\Delta_y,\Delta_z$
% & Grid cell dimensions along the $x$, $y$, and $z$ axes \\
% $H$ & UAV flight altitude \\
% $s,d$ & Departure and destination cells \\
% $\Nset_4(i,j)$ & Four-neighbor set of grid $(i,j)$ \\
% $\Kset$ & Set of BSs \\
% $K$ & Number of BSs \\
% $B$ & Common fixed bandwidth of each BS \\
% $\bm p_k$ & Location of BS $k$ \\
% $\mathcal M_k$ & CGM of BS $k$ \\
% $g_{k,i,j}^{(n)}$ & Locally averaged channel power gain \\
% $\eta_{k,i,j}$ & Spectral efficiency from BS $k$ to grid $(i,j)$ \\
% $A_{i,j}$ & Corridor-selection indicator \\
% $x_{k,i,j}$ & BS coverage indicator \\
% $u$ & Aerial traffic density in UAVs per grid cell \\
% $u_k$ & Aerial-user load of BS $k$ \\
% $R_0$ & Minimum per-UAV rate \\
% $\alpha_1,\alpha_2$ & Objective weights \\
% $z_{k,i,j}$ & Auxiliary variable for $ux_{k,i,j}$ \\
% \hline
% \end{tabular}
% \end{table}

\section{Joint Air Corridor and BS Coverage Planning}
\label{sec:problem}

Based on the CKM-assisted communication model developed in the previous section, we jointly design the air corridor and the BS coverage to enhance the aerial traffic density that can be supported by the terrestrial cellular network while minimizing the corridor length. In this section, we first represent the air corridor and specify its connectivity constraints, and then formulate the joint air corridor planning and BS coverage problem.

\subsection{Air Corridor Representation and Connectivity Constraints}

For each airspace grid cell $(i,j)\in\Nset^2$, define the binary corridor-selection variable $A_{i,j}\in\{0,1\}$, where $A_{i,j}=1$ indicates that grid cell $(i,j)$ is included in the air corridor, and $A_{i,j}=0$ otherwise.

The prescribed departure and destination cells must be included in the corridor, i.e.,
\begin{equation}
    A_{i_{\rm s},j_{\rm s}}=1,
    \quad
    A_{i_{\rm d},j_{\rm d}}=1.
    \label{eq:endpoint_selection}
\end{equation}
To form an air corridor connecting the two endpoints, the departure and destination cells are each required to have exactly one selected neighboring cell:
\begin{align}
    \sum_{(i',j')\in\Nset_4(i_{\rm s},j_{\rm s})}
    A_{i',j'}
    &=1,
    \label{eq:source_degree}\\
    \sum_{(i',j')\in\Nset_4(i_{\rm d},j_{\rm d})}
    A_{i',j'}
    &=1.
    \label{eq:destination_degree}
\end{align}
For every other selected grid cell, exactly two of its four-neighbor cells should also belong to the corridor. This requirement is imposed by

\begin{equation}
\begin{aligned}
    2A_{i,j}
    &\le
    \sum_{(i',j')\in\Nset_4(i,j)} A_{i',j'}
    \le
    2+2(1-A_{i,j}),\\
    &\hspace{-8mm}
    \forall (i,j)\in
    \Nset^2\setminus
    \left\{
        (i_{\rm s},j_{\rm s}),
        (i_{\rm d},j_{\rm d})
    \right\}.
\end{aligned}
\label{eq:internal_degree}
\end{equation}
When $A_{i,j}=1$, \eqref{eq:internal_degree} requires exactly two selected neighbors, whereas it does not impose a degree requirement on an unselected cell.

\subsection{Problem Formulation}

Based on the air corridor representation and the cellular communication model developed above, we jointly optimize the air corridor layout, the BS coverage of the selected grid cells, and the normalized aerial traffic density. First, each selected grid cell is covered by exactly one BS, whereas an unselected grid cell is not covered by any BS. Accordingly,
\begin{equation}
    \sum_{k=1}^{K}x_{k,i,j}
    =
    A_{i,j},
    \quad
    \forall (i,j)\in\Nset^2.
    \label{eq:association_constraint}
\end{equation}
Let $R_0$ denote the minimum rate requirement of each active UAV. Equation~\eqref{eq:per_uav_rate} estimates the rate based on the average BS load. A UAV can use at most the full bandwidth $B$, even when $u_k<1$. Thus, BS $k$ can serve cell $(i,j)$ only if $B\eta_{k,i,j}\ge R_0$. The load-based rate constraint is
\begin{equation}
    \sum_{k=1}^{K}
    \frac{x_{k,i,j}B\eta_{k,i,j}}{u_k}
    \ge
    R_0 A_{i,j},
    \quad
    \forall (i,j)\in\Nset^2.
    \label{eq:rate_constraint}
\end{equation}
In the rate and resource sums, unassigned links contribute zero and their fractions are evaluated only for assigned links.

With uniform grid resolution, corridor length is measured by the number of selected cells. Using weights $\alpha_1$ and $\alpha_2$, the joint planning problem is
\begin{subequations}
\label{prob:p1}
\begin{align}
\mathrm{(P1)}:&
\underset{
    \{A_{i,j}\},
    \{x_{k,i,j}\},
    u,
    \{u_k\}
}{\min}
\quad
\alpha_1
\sum_{i=1}^{N}\sum_{j=1}^{N}A_{i,j}
-\alpha_2 u
\label{eq:p1_obj}\\
\text{s.t.}\quad&
u_k
=
u
\sum_{i=1}^{N}\sum_{j=1}^{N}x_{k,i,j},
\forall k\in\Kset,
\label{eq:p1_load}\\
&
\sum_{k=1}^{K}
\frac{x_{k,i,j}B\eta_{k,i,j}}{u_k}
\ge
R_0 A_{i,j},
\quad
\forall (i,j)\in\Nset^2,
\label{eq:p1_rate}\\
&
\sum_{k=1}^{K}x_{k,i,j}
=
A_{i,j},
\forall(i,j)\in\Nset^2,
\label{eq:p1_assoc}\\
&
A_{i_{\rm s},j_{\rm s}}
=
A_{i_{\rm d},j_{\rm d}}
=
1,
\label{eq:p1_endpoints}\\
&
\sum_{(i',j')\in\Nset_4(i_{\rm s},j_{\rm s})}
A_{i',j'}
=
1,
\label{eq:p1_start}\\
&
\sum_{(i',j')\in\Nset_4(i_{\rm d},j_{\rm d})}
A_{i',j'}
=
1,
\label{eq:p1_destination}\\
&
2A_{i,j}
\le
\sum_{(i',j')\in\Nset_4(i,j)}A_{i',j'}
\le
2+2(1-A_{i,j}),
\notag\\[-1mm]
&
\hspace{7mm}
\forall(i,j)\in
\Nset^2\setminus
\left\{
(i_{\rm s},j_{\rm s}),
(i_{\rm d},j_{\rm d})
\right\},
\label{eq:p1_degree}\\
&
u_{\min}\le u\le u_{\max},
\label{eq:p1_u0}\\
&
x_{k,i,j}\in\{0,1\},
\forall k\in\Kset,\;
(i,j)\in\Nset^2,
\label{eq:p1_x}\\
&
A_{i,j}\in\{0,1\},
\forall(i,j)\in\Nset^2.
\label{eq:p1_A}
\end{align}
\end{subequations}
In \eqref{eq:p1_obj}, $\alpha_1>0$ and $\alpha_2>0$ weight the corridor length and the supported normalized aerial traffic density, respectively. Constraints \eqref{eq:p1_load} and \eqref{eq:p1_rate} couple the BS coverage with the aerial traffic load and the corresponding rate requirement, while \eqref{eq:p1_assoc} ensures that each selected grid cell is covered by exactly one BS. Constraints \eqref{eq:p1_endpoints}--\eqref{eq:p1_degree} specify the endpoint and degree conditions for the air corridor without introducing auxiliary flow variables. When $\alpha_1>0$, the corridor length term in the objective ensures that no isolated cycles are present in any optimal solution. The following theorem formalizes this property.

\begin{theorem}
\label{thm:connectivity}
For $\alpha_1>0$, the grid cells selected by every globally optimal
solution of (P1) form exactly one connected simple path from
$(i_{\rm s},j_{\rm s})$ to $(i_{\rm d},j_{\rm d})$. Hence,
\eqref{eq:p1_endpoints}--\eqref{eq:p1_degree}, together with the
objective in \eqref{eq:p1_obj}, exclude isolated selected cycles at
optimality.
\end{theorem}

\begin{proof}
See Appendix~\ref{app:connectivity}.
\end{proof}

Problem~(P1) involves discrete corridor and BS coverage
decisions together with a continuous traffic density
variable, subject to nonlinear rate constraints and
corridor connectivity requirements. The number of binary
variables grows with both the map size and the number of
BSs, leading to a large combinatorial search space and
making the computation of an optimal solution demanding.
To facilitate its solution, we first derive an exact MILP
reformulation. For large problem instances, we further
develop a scalable algorithm to reduce the computational
cost while maintaining good solution quality.

\section{Exact Solution to Problem (P1)}
\label{sec:exact}

In this section, we derive an exact MILP reformulation of
Problem~(P1) by equivalently reformulating the rate constraints
and exactly linearizing the products between $u$ and
$x_{k,i,j}$.

Using the unique BS coverage in \eqref{eq:p1_assoc}, the rate requirement
can be equivalently enforced by
\begin{equation}
    R_0u_k
    \le
    B\eta_{k,i,j}
    +
    M_{k,i,j}(1-x_{k,i,j}),
    \quad
    \forall k\in\Kset,\;
    (i,j)\in\Nset^2,
    \label{eq:big_m_rate}
\end{equation}
where
\begin{equation}
    M_{k,i,j}
    =
    \max\left\{
        0,\,
        R_0u_{\max}N^2-B\eta_{k,i,j}
    \right\}.
    \label{eq:big_m_value}
\end{equation}
The remaining nonlinear term in \eqref{eq:p1_load} is
$ux_{k,i,j}$. We introduce the auxiliary variable
\begin{equation}
    z_{k,i,j}=ux_{k,i,j},
    \quad
    \forall k\in\Kset,\;
    (i,j)\in\Nset^2.
    \label{eq:z_definition}
\end{equation}
Since $x_{k,i,j}$ is binary and
$u_{\min}\le u\le u_{\max}$, this product is represented exactly by
\begin{subequations}
\label{eq:product_linearization}
\begin{align}
    z_{k,i,j}
    &\le u_{\max}x_{k,i,j},
    \label{eq:lin_z1}\\
    z_{k,i,j}
    &\ge u_{\min}x_{k,i,j},
    \label{eq:lin_z2}\\
    z_{k,i,j}
    &\le u-u_{\min}(1-x_{k,i,j}),
    \label{eq:lin_z3}\\
    z_{k,i,j}
    &\ge u-u_{\max}(1-x_{k,i,j}),
    \label{eq:lin_z4}
\end{align}
\end{subequations}
for all $k\in\Kset$ and $(i,j)\in\Nset^2$. Accordingly,
\eqref{eq:p1_load} becomes
\begin{equation}
    u_k
    =
    \sum_{i=1}^{N}\sum_{j=1}^{N}
    z_{k,i,j},
    \quad
    \forall k\in\Kset.
    \label{eq:linear_load}
\end{equation}

Problem~(P1) can then be equivalently reformulated as
\begin{subequations}
\label{prob:p2}
\begin{align}
\mathrm{(P2)}:\quad&
\underset{
    \substack{
        \{A_{i,j}\},\,\{x_{k,i,j}\} \\
        \{z_{k,i,j}\},\,u,\{u_k\} 
    }
}{\min}
\quad
\alpha_1
\sum_{i=1}^{N}\sum_{j=1}^{N}A_{i,j}
-\alpha_2u
\label{eq:p2_obj}\\
\text{s.t.}\quad
&
u_k
=
\sum_{i=1}^{N}\sum_{j=1}^{N}z_{k,i,j},
\forall k\in\Kset,
\label{eq:p2_load}\\
&
R_0u_k
\le
B\eta_{k,i,j}
+
M_{k,i,j}(1-x_{k,i,j}),\label{eq:p2_rate}\\
&
\forall k\in\Kset,\;
(i,j)\in\Nset^2,
\notag\\
&
\eqref{eq:product_linearization},
\eqref{eq:p1_assoc}
-\eqref{eq:p1_A}.
\label{eq:p2_original}
\end{align}
\end{subequations}
The reformulations in \eqref{eq:big_m_rate} and
\eqref{eq:product_linearization} are exact for binary $x_{k,i,j}$; hence,
(P2) is an exact MILP reformulation of (P1). Moreover, since $\alpha_1>0$, Theorem~\ref{thm:connectivity}
ensures that an optimal solution of (P2) yields a connected
air corridor without additional connectivity recovery.

Problem~(P2) contains $(K+1)N^2$ binary variables and $KN^2+K+1$ continuous variables. For small-scale maps, this exact MILP reformulation can be directly solved by standard solvers to obtain the globally optimal solution of (P1). As the map resolution and the number of BSs increase, however, the number of binary variables grows as $\mathcal O(KN^2)$, leading to a rapidly increasing computational burden for exact MILP optimization. Directly solving the full-map formulation therefore becomes impractical for large-scale instances. To address this scalability issue, we develop a computationally efficient heuristic algorithm in the next section, which combines polynomial-time path search with corridor-restricted coverage recovery.

\section{Scalable Suboptimal Optimization for Large-Scale Maps}
\label{sec:mrcsp_scalable}

Although Problem~(P2) provides an exact MILP reformulation, its complexity grows with all grid cells and BS coverage variables. We
therefore develop a scalable two-stage solution. First, for each trial traffic density, we relax the original rate constraints into additive BS resource constraints, yielding a multi-resource constrained shortest path (MRCSP) problem, which is further handled by a Lagrangian-relaxation-based A* search to generate candidate corridors.
Second, with each distinct candidate corridor fixed, we
solve the MILP~(P4) to jointly optimize BS coverage and
the continuous traffic density under the original model
with equal resource sharing.

\subsection{Additive MRCSP Relaxation}

In this section, we first derive an additive MRCSP relaxation for generating
candidate corridors at each trial traffic density. In the original problem, directly
embedding $u$ into a shortest-path search is difficult because the term \(uR_0/(B\eta_{k,i,j})\) associated with assigning grid cell \((i,j)\) to BS \(k\) depends on \(u\). In particular, the resource
consumption of serving grid cell $(i,j)$ with BS $k$ scales with
$uR_0/(B\eta_{k,i,j})$ and therefore cannot be assigned a fixed resource
weight while $u$ remains an optimization variable.

To obtain fixed resource coefficients during corridor generation, we
introduce a set of trial traffic densities
\begin{equation}
    u_{\min}
    =
    u^{(1)}
    <
    \cdots
    <
    u^{(V)}
    =
    u_{\max}.
    \label{eq:traffic_levels}
\end{equation}
For each trial level \(u^{(v)}\), we temporarily fix \(u=u^{(v)}\), so that \(u^{(v)}R_0/(B\eta_{k,i,j})\) becomes fixed for every \(k\) and \((i,j)\). Since
$-\alpha_2u^{(v)}$ is also constant, the path-generation problem
focuses on finding a short corridor while satisfying the communication requirements. 

Consider an arbitrary candidate corridor $\mathcal P$ together with
the BS coverage variables
$\{x_{k,i,j}\}_{k\in\mathcal K,(i,j)\in\mathcal P}$. If BS $k$ serves at least one grid cell on
$\mathcal P$, its aggregate UAV load is proportional to the number of
corridor cells covered by it. Under $u=u^{(v)}$, the original
per-UAV rate requirement is therefore equivalent to
\begin{equation}
    u^{(v)}R_0
    \sum_{(i,j)\in\mathcal P}x_{k,i,j}
    \le
    B\min_{\substack{(i,j)\in\mathcal P\\x_{k,i,j}=1}}
    \eta_{k,i,j},
    \label{eq:original_fixed_u_constraint}
\end{equation}
for every BS $k$ serving at least one corridor cell. Equation
\eqref{eq:original_fixed_u_constraint} shows that the admissible load of BS \(k\) is determined by the minimum spectral efficiency \(\eta_{k,i,j}\) among the corridor cells covered by BS \(k\), with \(B\) fixed. The constraint in \eqref{eq:original_fixed_u_constraint} is
non-additive across corridor cells because all cells covered by the
same BS are coupled through their common aggregate load. To obtain an
additive representation suitable for resource constrained path search,
observe that if a corridor \(\mathcal P\) and its BS coverage \(\{x_{k,i,j}\}\) satisfy \eqref{eq:original_fixed_u_constraint}, they also satisfy 
\begin{equation}
    \sum_{(i,j)\in\mathcal P}
    x_{k,i,j}
    \frac{u^{(v)}R_0}{B\eta_{k,i,j}}
    \le
    \frac{
        u^{(v)}R_0
        \sum_{(i,j)\in\mathcal P}x_{k,i,j}
    }{
        \displaystyle
        B\min_{\substack{(i,j)\in\mathcal P\\x_{k,i,j}=1}}
        \eta_{k,i,j}
    }\le 1.
    \label{eq:additive_bound}
\end{equation}
Therefore, we relax the original equal-sharing rate requirement by
adopting the following BS-indexed additive constraints
\begin{equation}
    \sum_{(i,j)\in\mathcal P}
    x_{k,i,j}
    \frac{u^{(v)}R_0}{B\eta_{k,i,j}}
    \le 1,
    \quad
    \forall k\in\Kset.
    \label{eq:additive_surrogate}
\end{equation}
The resulting fixed-density path-generation problem is
\begin{subequations}
\label{prob:mrcsp_v}
\begin{align}
(\mathrm{P3}\text{-}v):&\quad
\min_{\mathcal P,\{x_{k,i,j}\}}
\quad
 |\mathcal P|
\\
\mathrm{s.t.}\quad
&
\sum_{(i,j)\in\mathcal P}
x_{k,i,j}
\frac{u^{(v)}R_0}{B\eta_{k,i,j}}
\le 1,
\quad
\forall k\in\Kset,
\\
&
\sum_{k=1}^{K}x_{k,i,j}=1,
\quad
\forall (i,j)\in\mathcal P,
\\
&
x_{k,i,j}\in\{0,1\},
\quad
\forall k\in\Kset, (i,j)\in\mathcal P.
\end{align}
\end{subequations}

Problem~(P3-$v$) can be viewed as an MRCSP problem, with one constraint dimension associated with each BS. For a fixed trial density \(u^{(v)}\), assigning grid cell \((i,j)\) to BS \(k\) contributes
$u^{(v)}R_0/(B\eta_{k,i,j})$ to the \(k\)-th constraint and zero to the constraints associated with the other BSs. Each selected grid cell also contributes one unit to the path length. Therefore, the corridor and its BS coverage can be jointly determined within the same path search.

\begin{table}[t]
\centering
\caption{Key Notations for the Large-Scale Solution}
\label{tab:large_scale_notation}
\footnotesize
\renewcommand{\arraystretch}{1.1}
\setlength{\tabcolsep}{4pt}
\begin{tabular}{p{0.18\columnwidth} p{0.71\columnwidth}}
\hline
Notation & Meaning \\
\hline
$u$
& Traffic density in the original problem \\

$u^{(v)}$
& $v$-th trial traffic level used for path generation \\

$V$
& Number of trial traffic levels \\

$\mathcal P^{(v,t)}$
& Corridor generated at level $v$ and iteration $t$ \\

$\lambda_k^{(t)}$
& Lagrangian multiplier of  BS $k$ \\

$\phi_v(\boldsymbol{\lambda})$
& Lagrangian dual function at traffic level $u^{(v)}$ \\

$\boldsymbol{e}^{(v,t)}$
& BS resource constraint residual vector \\

$\beta_0$
& Multiplier step-size coefficient \\

$T$
& Maximum number of iterations per traffic level \\

$u^\star(\mathcal P)$
& Optimized traffic density of corridor $\mathcal P$\\

$J(\mathcal P)$
& Original objective value of the recovered corridor $\mathcal P$ \\
\hline
\end{tabular}
\end{table}

\subsection{Lagrangian Relaxation and A*-Based Path Search}

To generate candidate corridors for the MRCSP in
Problem~(P3-$v$), we apply Lagrangian relaxation to its
$K$ BS resource constraints~\cite{handler1980dual},
incorporating BS resource consumption into the objective
through Lagrangian multipliers. The resulting subproblem reduces to a shortest path problem, which we solve using A*. This algorithm prioritizes grid cells
according to the sum of the accumulated path cost and
a heuristic estimate of the remaining cost to the
destination. The multipliers
are then updated based on the resulting BS resource usage,
and the A* search is repeated under the adjusted resource
penalties to generate candidate corridors.

Before applying Lagrangian relaxation, we retain a necessary
feasibility condition implied by Problem~(P3-$v$). If cell $(i,j)$ is
covered by BS $k$, then
$u^{(v)}R_0/(B\eta_{k,i,j})\le 1$ must hold. We therefore define
\begin{equation}
    \Kset_{i,j}^{(v)}
    \triangleq
    \left\{
        k\in\Kset:
        B\eta_{k,i,j}\ge u^{(v)}R_0
    \right\}.
    \label{eq:feasible_bs_set}
\end{equation}
Restricting the BSs allowed to cover cell $(i,j)$ to
$\Kset_{i,j}^{(v)}$ does not change the feasible set of
Problem~(P3-$v$), while preventing the subsequent relaxed search from
selecting a BS that cannot serve the cell at the trial density
even when serving that cell alone.

We next apply Lagrangian relaxation to the $K$ additive BS constraints
in (P3-$v$). Let $\lambda_k\ge0$ denote the multiplier associated with
the constraint of BS $k$. The Lagrangian dual function
for the trial density $u^{(v)}$ is
\begin{equation}
\begin{aligned}
\phi_v(\boldsymbol{\lambda})
&=\min_{\mathcal P,\{x_{k,i,j}\}}\\
&\quad\Bigg\{|\mathcal P|
+\sum_{k=1}^{K}\lambda_k\left(
\sum_{(i,j)\in\mathcal P}x_{k,i,j}
\frac{u^{(v)}R_0}{B\eta_{k,i,j}}-1\right)\Bigg\},
\end{aligned}
\label{eq:lagrangian_dual_function}
\end{equation}
where the corridor constraints and the coverage constraints are
retained, and cell $(i,j)$ can only be covered by a BS in
$\Kset_{i,j}^{(v)}$. The corresponding dual problem is
\begin{equation}
\max_{\boldsymbol{\lambda}\succeq\mathbf 0}
\phi_v(\boldsymbol{\lambda}).
\label{prob:lagrangian_dual}
\end{equation}

For a given
$\boldsymbol{\lambda}$, the serving BS can be chosen independently for each selected grid cell. Specifically, if grid cell \((i,j)\) is included in the corridor, its provisional serving BS is chosen as
\begin{equation}
k_{i,j}^{\star}(\boldsymbol{\lambda})
\in
\arg\min_{k\in\Kset_{i,j}^{(v)}}
\left\{
\lambda_k
\frac{u^{(v)}R_0}{B\eta_{k,i,j}}
\right\}.
\label{eq:provisional_bs}
\end{equation}
Accordingly, including grid cell \((i,j)\) contributes
\begin{equation}
1+
\lambda_{k_{i,j}^{\star}}
\frac{u^{(v)}R_0}
{B\eta_{k_{i,j}^{\star},i,j}}
\label{eq:lagrangian_cell_cost}
\end{equation}
to the path-dependent part of the Lagrangian objective. The first term in \eqref{eq:lagrangian_cell_cost} accounts for the contribution of the selected cell to
the corridor length, while the second term is the Lagrangian penalty
associated with the BS resources consumed by serving that cell. We run A* on the airspace
grid using the grid cell costs in
\eqref{eq:lagrangian_cell_cost} to obtain a candidate
corridor connecting the departure and destination cells.
% The Manhattan distance
% $|i-i_{\rm d}|+|j-j_{\rm d}|$ is used as the heuristic.
% Since each grid cell cost is at least one, this heuristic
% does not overestimate the remaining path cost.
% Hence, A* finds a path of minimum total cost and solves
% the inner minimization exactly for the given multipliers.

To solve the dual problem in \eqref{prob:lagrangian_dual}, we use
projected supergradient ascent. At iteration $t$, A* solves the inner
minimization for $\boldsymbol{\lambda}^{(t)}$, yielding the corridor
$\mathcal P^{(v,t)}$ and the corresponding provisional BS coverage
$\{x_{k,i,j}^{(v,t)}\}$ determined by \eqref{eq:provisional_bs}.
Because the inner minimization is solved optimally, the residuals of
the relaxed resource constraints form a supergradient of the dual
function at $\boldsymbol{\lambda}^{(t)}$. Its $k$-th component is
\begin{align}
e_k^{(v,t)}
=
\sum_{(i,j)\in\mathcal P^{(v,t)}}
x_{k,i,j}^{(v,t)}
\frac{u^{(v)}R_0}{B\eta_{k,i,j}}
-1,
\quad
\forall k\in\Kset.
\label{eq:resource_residual}
\end{align}
Here, $e_k^{(v,t)}>0$ indicates that the resource constraint of BS $k$
is violated, whereas $e_k^{(v,t)}<0$ indicates remaining resource.

The multipliers are then updated as
\begin{align}
\lambda_k^{(t+1)}
=
\left[
\lambda_k^{(t)}
+
\frac{\beta_0}{\sqrt{t+1}}
\frac{e_k^{(v,t)}}{
\max\!\left\{1,\|\boldsymbol e^{(v,t)}\|_2\right\}}
\right]^+,
\quad
\forall k\in\Kset,
\label{eq:multiplier_update}
\end{align}
where $\beta_0>0$ is a step-size parameter and
$[x]^+=\max\{x,0\}$. A positive residual increases the corresponding
multiplier, making coverage by that BS more costly in the next
A* search. Consequently, the provisional BS coverage and the resulting
corridor may change in the next iteration. The normalization bounds the norm of the update direction by one
without amplifying small residuals, thereby avoiding excessively large
changes in the multipliers. The factor $1/\sqrt{t+1}$ further reduces
the step size as the iterations proceed. The A* search and multiplier update are repeated for $T$
iterations at each trial traffic level.

\begin{algorithm}[t]
\caption{Lagrangian--A* Corridor Search and Recovery}
\label{alg:lagrangian_astar}
\begin{algorithmic}[1]
\Require CKM-derived spectral efficiencies $\{\eta_{k,i,j}\}$,
bandwidth $B$, traffic levels
$\{u^{(v)}\}_{v=1}^{V}$, maximum multiplier iterations $T$,
and step-size factor $\beta_0$

\State $\mathcal S\gets\varnothing$ \Comment{Set of candidate corridors}

\For{$v=1,\ldots,V$}
    \State Construct the admissible BS sets
    $\{\Kset_{i,j}^{(v)}\}$
    \State $\boldsymbol{\lambda}^{(0)}\gets\mathbf 0$

    \For{$t=0,\ldots,T-1$}
        \State Compute the Lagrangian grid-cell costs from
        \eqref{eq:lagrangian_cell_cost}

        \State Run A* with the Manhattan-distance heuristic to obtain
        $\mathcal P^{(v,t)}$ and its provisional BS coverage
        $\{x_{k,i,j}^{(v,t)}\}$

        \State $\mathcal S\gets
        \mathcal S\cup\{\mathcal P^{(v,t)}\}$

        \State Compute $\boldsymbol e^{(v,t)}$ from
        \eqref{eq:resource_residual}

        \State Update $\boldsymbol{\lambda}^{(t+1)}$ according to
        \eqref{eq:multiplier_update}
    \EndFor
\EndFor

\State $J^{\mathrm{best}}\gets+\infty$,
$\mathcal P^{\mathrm{best}}\gets\varnothing$

\ForAll{$\mathcal P\in\mathcal S$}
    \State Solve (P4) for $\mathcal P$

    \If{(P4) is feasible}
        \State Compute $J(\mathcal P)$ from
        \eqref{eq:recovered_objective}

        \If{$J(\mathcal P)<J^{\mathrm{best}}$}
            \State $J^{\mathrm{best}}\gets J(\mathcal P)$
            \State $\mathcal P^{\mathrm{best}}\gets\mathcal P$
            \State Store its optimized BS coverage and
            $u^\star(\mathcal P)$
        \EndIf
    \EndIf
\EndFor

\State \Return $\mathcal P^{\mathrm{best}}$, its optimized BS
coverage, and $u^\star(\mathcal P^{\mathrm{best}})$

\end{algorithmic}
\end{algorithm}

\subsection{Feasible-Solution Recovery}

The provisional BS coverage generated by a Lagrangian subproblem
need not satisfy all BS resource constraints. Therefore, each distinct
candidate corridor $\mathcal P$ is evaluated by reoptimizing its BS
coverage and traffic density. Specifically, the optimization problem becomes
\begin{subequations}
\label{prob:recovery}
\begin{align}
(\mathrm{P}4):\quad&
\underset{
    \substack{
        \{x_{k,i,j}\},\,\{z_{k,i,j}\},\,u,\,\{u_k\}
    }
}{\max}
\quad u
\label{eq:recovery_obj}\\
\mathrm{s.t.}\quad
&
u_k
=
\sum_{(i,j)\in\mathcal P}z_{k,i,j},
\quad
\forall k\in\Kset,
\label{eq:recovery_load}\\
&
R_0u_k
\le
B\eta_{k,i,j}+M_{k,i,j}(1-x_{k,i,j}),
\notag\\[-1mm]
&\hspace{27mm}
\forall k\in\Kset,\;(i,j)\in\mathcal P,
\label{eq:recovery_rate}\\
&
\sum_{k=1}^{K}x_{k,i,j}=1,
\quad
\forall(i,j)\in\mathcal P,
\label{eq:recovery_assoc}\\
&
u_{\min}\le u\le u_{\max},
\label{eq:recovery_density}\\
&
\eqref{eq:product_linearization},
\quad \forall k\in\Kset,\;(i,j)\in\mathcal P,
\label{eq:recovery_linearization}\\
&
x_{k,i,j}\in\{0,1\},
\quad
\forall k\in\Kset,\;(i,j)\in\mathcal P.
\label{eq:recovery_binary}
\end{align}
\end{subequations}
Problem~(P4) is a MILP
with $K|\mathcal P|$ binary coverage variables and can be
solved using the standard MILP solvers. The recovered objective value of corridor
$\mathcal P$ is
\begin{equation}
J(\mathcal P)
=
\alpha_1|\mathcal P|-\alpha_2u^\star(\mathcal P).
\label{eq:recovered_objective}
\end{equation}
The overall procedure is summarized in
Algorithm~\ref{alg:lagrangian_astar}.

\subsection{Convergence of the Lagrangian Dual Search}
\label{subsec:rcsp_theory}

For each trial traffic density $u^{(v)}$, the Lagrangian--A* search
generates a sequence of dual values through the multiplier update in
\eqref{eq:multiplier_update}. Define the best dual value obtained after
$T$ iterations as
\begin{equation}
    D_v^{(T)}
    \triangleq
    \max_{0\le t<T}
    \phi_v\!\left(\boldsymbol\lambda^{(t)}\right),
    \label{eq:best_dual_bound}
\end{equation}
and let $ \phi_v^\star\triangleq\max_{\boldsymbol\lambda\succeq\bm 0}\phi_v(\boldsymbol\lambda)$
denote the optimal value of the corresponding Lagrangian dual problem.

\begin{theorem}
\label{thm:dual_convergence}
Suppose that (P3-$v$) is feasible and that each Lagrangian subproblem
is solved exactly. Then,
\begin{equation}
    \lim_{T\rightarrow\infty}D_v^{(T)}
    =
    \phi_v^\star.
    \label{eq:dual_convergence}
\end{equation}
Moreover, for a fixed problem instance and step-size parameter
$\beta_0$, the best-dual-value gap satisfies
\begin{equation}
    \phi_v^\star-D_v^{(T)}
    =
    O\!\left(
        \frac{\log(T+1)}{\sqrt{T}}
    \right).
    \label{eq:dual_convergence_rate}
\end{equation}
\end{theorem}

\begin{proof}
See Appendix~\ref{app:dual_convergence}.
\end{proof}

Theorem~\ref{thm:dual_convergence} shows that the normalized projected
supergradient update asymptotically attains the best lower bound provided
by the corresponding Lagrangian dual problem. In the proposed method,
each Lagrangian subproblem reduces to an ordinary shortest-path problem
and is solved optimally by A*.

\subsection{Complexity Analysis}

For each trial traffic level \(v\) and multiplier iteration \(t\), evaluating
\eqref{eq:lagrangian_cell_cost} over all grid cells requires
$O(KN^2)$ operations. The four-neighbor grid contains $N^2$ vertices
and $O(N^2)$ edges, and one A* search has worst-case complexity
$O(N^2\log N)$. Hence, the total complexity of the Lagrangian--A*
path-search stage is $O\!\left(VT N^2\left(K+\log N\right)\right)$.

Each distinct corridor generated during the search is subsequently
evaluated by (P4). For a corridor containing
$L$ cells, the recovery problem contains $KL$ binary coverage
variables and its formulation size depends on the corridor length
rather than on all $N^2$ grid cells. Therefore, the large-scale
geometric search is performed entirely by the polynomial-time A*
procedure, while integer optimization is restricted to the generated
one-dimensional corridors.

% Queue the scene overview for the opening page of Numerical Results.
\begin{figure*}[!t]
    \centering
    \includegraphics[width=\textwidth]{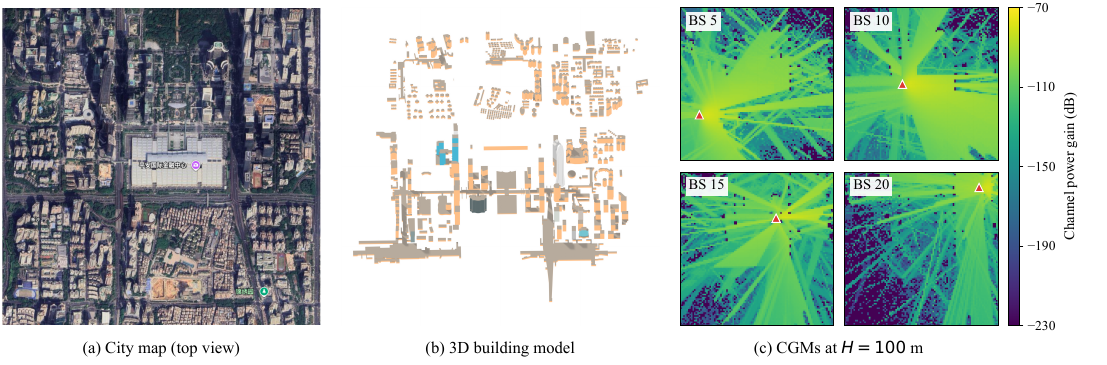}
    \caption{Urban scene and representative CGMs: (a) top-view city map;
    (b) reconstructed 3D building model; and (c) CGMs at $H=100$ m for BSs 5,
    10, 15, and 20. Red triangles
    mark the horizontal BS locations. }
    \label{fig:futian_scene_cgms}
\end{figure*}

\section{Numerical Results}

This section provides numerical results to evaluate the proposed joint
air corridor planning and BS coverage design. We consider a
horizontal low-altitude flight layer at $H=100$ m over urban environments discretized into square grids. The location-dependent
channel gains are provided by CKMs generated offline over the
corresponding 3D urban scenes using the RadioMapSolver in Sionna RT
v1.2.1~\cite{sionna}. The ray-tracing model accounts for line-of-sight
propagation, specular and diffuse reflections, and diffraction. A vertically polarized single-element antenna
with the radiation pattern specified in the 3rd Generation Partnership
Project (3GPP) Technical Report~38.901 is adopted at each BS
\cite{tr38901antenna}. The
resulting path-gain maps are sampled over the prescribed flight layer
according to the grid resolution of each scenario. The common simulation parameters are summarized in Table~\ref{tab:simulation_parameters}.

\begin{table}[t]
\centering
\caption{Simulation Parameters}
\label{tab:simulation_parameters}
\footnotesize
\renewcommand{\arraystretch}{1.08}
\begin{tabular}{l c}
\hline
\textbf{Parameter} & \textbf{Value} \\
\hline
Flight altitude, $H$ & $100$ m~\cite{tarafder2025digitaltwin} \\
Carrier frequency, $f_c$ & $3.5$ GHz~\cite{tarafder2025digitaltwin} \\
Ray samples per BS & $10^7$ \\
Maximum propagation depth & $3$ \\
System bandwidth, $B$ & $200$ MHz \\
Transmit gain, $G$ &$10$ dB \\
BS transmit power, $P$ & $1$ W~\cite{maeng2024strongest} \\
Receiver noise power, $N_0B$ & $-91$ dBm \\
Minimum per-UAV rate, $R_0$ & $10$ Mbit/s \\
Traffic density range, $[u_{\min},u_{\max}]$ & $[0.01,5]$ \\
Corridor length weight, $\alpha_1$ & $1$ \\
Traffic density weight, $\alpha_2$ & $100$ \\
\hline
\end{tabular}
\end{table}

\subsection{Benchmark Schemes and Evaluation Protocol}

\begin{table}[t]
\centering
\caption{Scenario Configurations for Numerical Evaluation}
\label{tab:scenario_configurations}
\footnotesize
\renewcommand{\arraystretch}{1.1}
\setlength{\tabcolsep}{2pt}
\begin{tabular}{@{}c c c c c c c@{}}
\hline
Scenario
& \shortstack{Area dimensions\\(m)}
& \shortstack{Map\\size}
& \shortstack{$\Delta$\\(m)}
& $K$
& $s=(i_{\rm s},j_{\rm s})$
& $d=(i_{\rm d},j_{\rm d})$ \\
\hline
Instance 0
& $2000\times2000$
& $100\times100$
& 20
& 20
& $(1,1)$
& $(100,100)$ \\
Instance 1
& $500\times500$
& $100\times100$
& 5
& 6
& $(4,51)$
& $(97,51)$ \\
Instance 2
& $500\times500$
& $100\times100$
& 5
& 6
& $(1,1)$
& $(100,100)$ \\
Instance 3
& $500\times500$
& $100\times100$
& 5
& 10
& $(4,50)$
& $(94,54)$ \\
Instance 4
& $10{,}000\times10{,}000$
& $200\times200$
& 50
& 20
& $(6,100)$
& $(195,105)$ \\
\hline
\end{tabular}
\end{table}

% Queue the wide corridor panel before its discussion.
\begin{figure*}[!t]
    \centering
    \includegraphics[width=\textwidth]{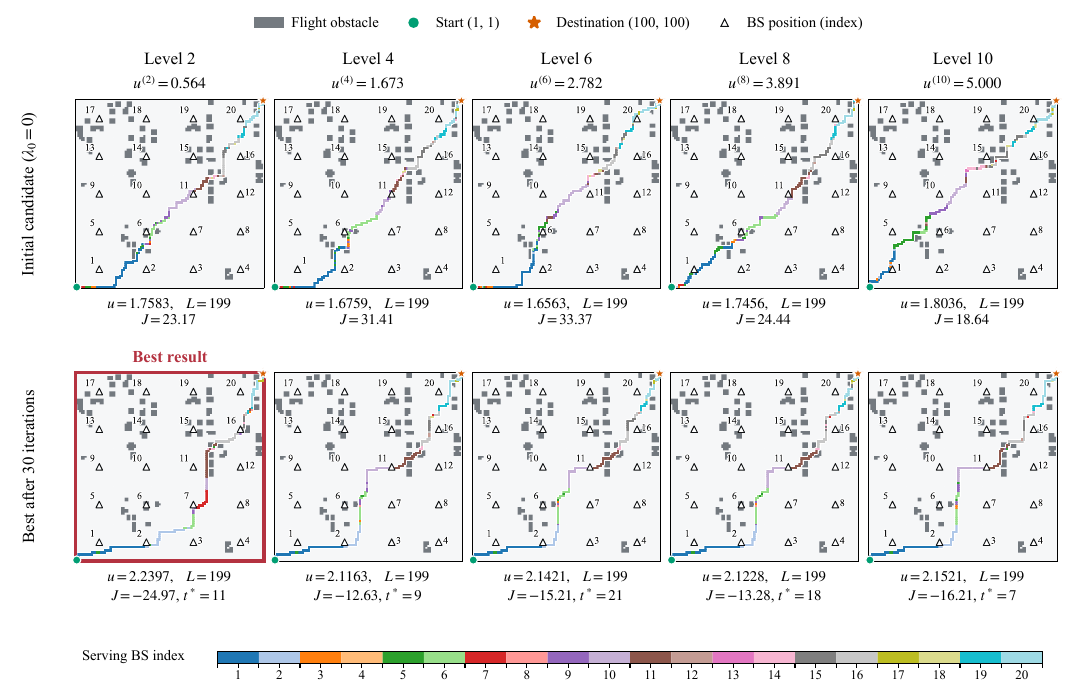}
    \caption{Initial and iteratively selected corridors at trial
density levels $v\in\{2,4,6,8,10\}$. The top row shows the first candidate
generated with zero initial multipliers. The bottom row shows the best
corridor found within 30 searches at each level;
$t^\star$ denotes its first appearance in the search sequence.
The red frame marks the lowest displayed objective value. }
    \label{fig:futian_paths_by_search_level}
\end{figure*}

To evaluate the proposed Lagrangian--A* method, we compare it with four benchmark schemes, which are described below. The first two benchmarks are based on shortest path enumeration, whereas the other two employ general heuristic search methods.

\begin{itemize}

    \item \emph{Shortest Paths with Strongest-BS Coverage
    (SP--SC):}
    Each corridor cell is served by the BS providing the highest spectral efficiency $\eta_{k,i,j}$, thereby determining the BS coverage along the corridor. The supported traffic density is then optimized under this fixed coverage design, and the candidate with the smallest original objective value is retained.

    \item \emph{Shortest Paths with Optimized Coverage
    (SP--OC):}
    For each candidate path, the BS coverage and traffic density are jointly optimized by solving Problem~(P4), and the candidate with the smallest original objective value is retained.

    \item \emph{Large Neighborhood Search (LNS):}
    Starting from a feasible corridor, LNS iteratively reoptimizes
    randomly selected local corridor neighborhoods using a restricted
    MILP while keeping the remaining corridor fixed. The resulting corridor replaces the current one only if it has a lower original objective value.
    \item \emph{Genetic Algorithm (GA):}
    Each chromosome represents a feasible corridor, which is evolved
    through selection, crossover, and local mutation. Each distinct
    candidate corridor is evaluated by solving Problem~(P4), and its
    fitness is determined by the resulting original objective.

\end{itemize}

The proposed method uses 10 uniformly spaced trial densities over
$[0.01,5]$ and 30 A* searches at each level. SP--SC and SP--OC evaluate the same 300 geometric candidates. GA uses a population of 16, two elites, a crossover probability of 0.8, and a mutation probability of 0.4. Unless otherwise stated, these settings apply to all instances, with an overall time limit of 1200 s per run.

\subsection{Performance Evaluation in an Urban Scenario}
\label{subsec:futian_trial_density}

We first consider Instance~0, which covers a
$2~\mathrm{km}\times2~\mathrm{km}$ urban area, to illustrate the operation and performance of the proposed
method. The geographic data are obtained from OpenStreetMap, and the building footprints and urban layout are used to reconstruct the corresponding 3D urban scene. Fig.~\ref{fig:futian_scene_cgms} presents the city map, the reconstructed building model, and representative CGMs for four BSs.

Fig.~\ref{fig:futian_paths_by_search_level} illustrates the corridor exploration at different trial traffic levels and multiplier iterations for \(R_0=10\) Mbit/s. Each trial level defines a different path search problem through the admissible BS sets and the values of \(u^{(v)}R_0/(B\eta_{k,i,j})\) for the candidate BS assignments at each grid cell. As the multipliers are updated, these differences result in different search costs and lead the algorithm to explore distinct corridor layouts and BS coverage patterns. Although all the displayed corridors connect \((1,1)\) and \((100,100)\) and contain 199 grid cells, their recovered traffic densities differ because they traverse regions with different channel conditions and distribute the traffic differently among the BSs. Each generated corridor is subsequently evaluated by solving Problem (P4), which optimizes the BS coverage and the supported traffic density under the original rate constraints. Among the displayed results, the best solution is generated at \(v=2\), which contains 199 grid cells and supports a traffic density of \(u=2.2397\).

\begin{figure}[!tb]
    \centering
    \includegraphics[width=0.8\columnwidth]{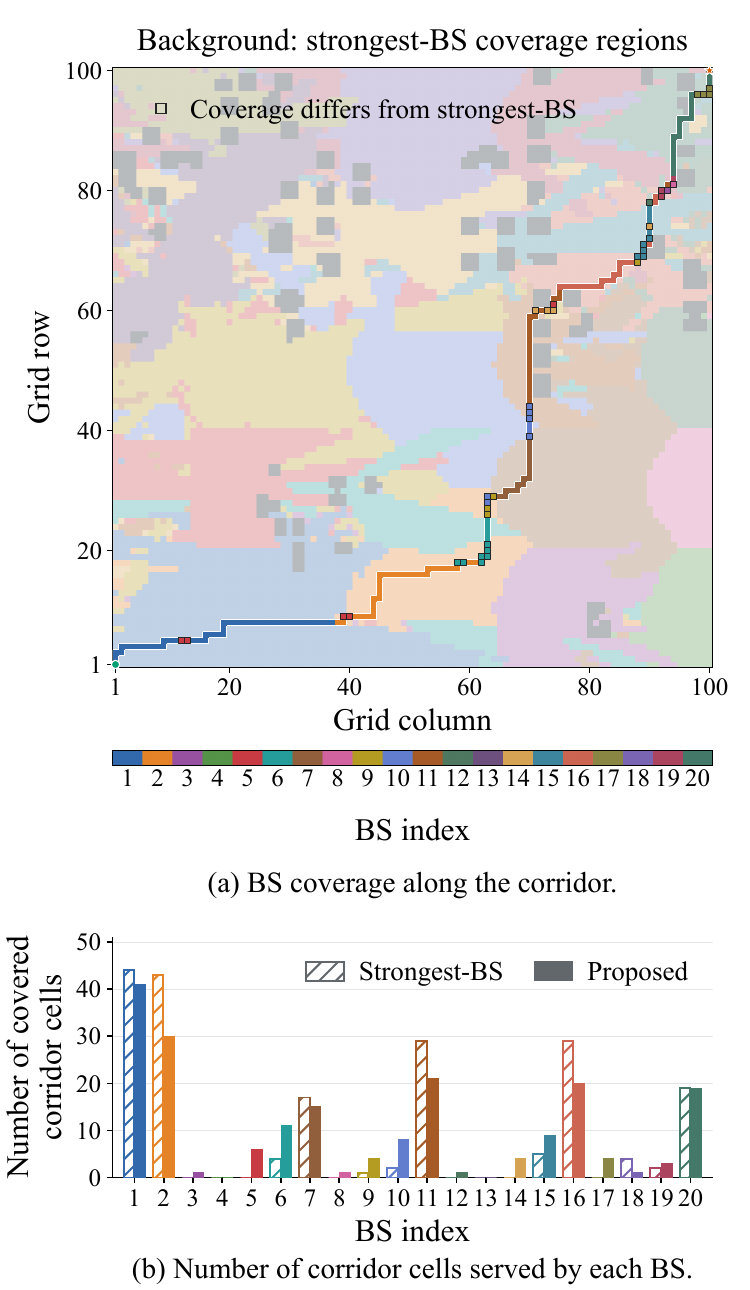}
    \caption{BS coverage comparison along the selected corridor.}
    \label{fig:futian_corridor_and_association}
\end{figure}

Fig.~\ref{fig:futian_corridor_and_association} further examines how the BS coverage is adjusted on the corridor selected in Fig.~\ref{fig:futian_paths_by_search_level}. The background in Fig.~\ref{fig:futian_corridor_and_association}(a) indicates the BS providing the highest spectral efficiency at each grid cell, whereas the corridor colors indicate the serving BS for each cell under the coverage design obtained by solving Problem (P4). Although many corridor cells are served by their strongest BS, the optimized solution deliberately assigns some cells to alternative BSs. As shown in Fig.~\ref{fig:futian_corridor_and_association}(b), these changes reduce the concentration of corridor cells on several heavily used BSs and redistribute part of the traffic to other BSs. This observation illustrates an important consequence of resource sharing. Selecting the strongest BS independently at each location does not necessarily maximize the supported traffic density, since the achievable rate also depends on how many corridor cells share the resources of the same BS. The coverage therefore needs to account for both link quality and the resulting BS load.

\subsection{Performance Comparison with Benchmark Schemes}

We next evaluate the solution quality and computational efficiency of the proposed method on the additional instances summarized in Table~\ref{tab:scenario_configurations}. These instances contain different map sizes, BS deployments, endpoint locations, obstacle layouts, and channel conditions. Instances 1--3 use \(100\times100\) grids, while Instance 4 uses a \(200\times200\) grid to examine the behavior of the algorithms on a larger map. We first use progressively larger cropped regions of Instance 1 to compare the proposed method with the full MILP formulation as the problem size increases.

% Figure sources and timing provenance: figure/instance1_small_map_comparison_sources.json.
% Draft data note: the first three MILP time values are user-edited plotting
% values and differ from the original logs; they remain pending verification.
% Proposed times are verified medians of 10 fixed-seed runs after warmup.
\begin{figure}[!tb]
    \centering
    \begin{subfigure}[t]{\columnwidth}
        \centering
        \includegraphics[width=0.9\linewidth]{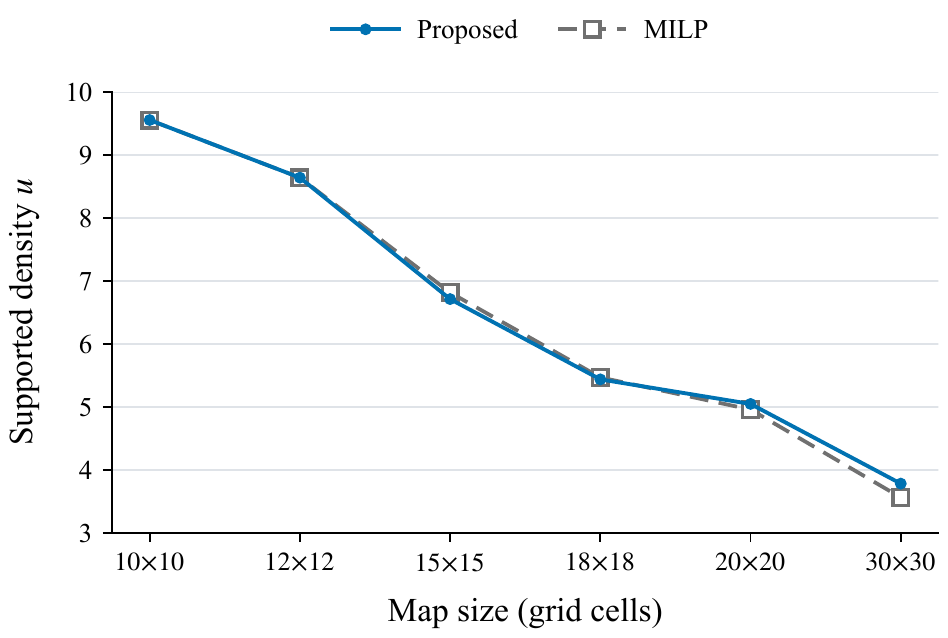}
        \caption{Supported density.}
        \label{fig:instance1_small_map_density}
    \end{subfigure}\par\vspace{0.8em}
    \begin{subfigure}[t]{0.9\columnwidth}
        \centering
        \includegraphics[width=\linewidth]{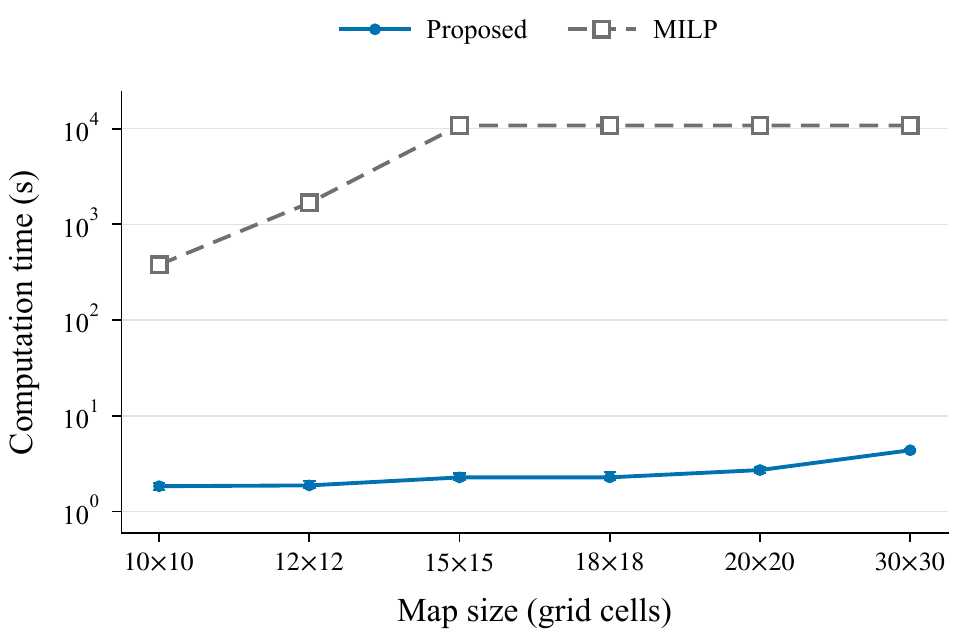}
        \caption{Computation time.}
        \label{fig:instance1_small_map_time}
    \end{subfigure}
    \caption{Performance comparison of the proposed method and MILP for different map sizes.}
    \label{fig:instance1_small_map_comparison}
\end{figure}

Fig.~\ref{fig:instance1_small_map_comparison} compares the proposed method with the full MILP formulation on progressively larger cropped maps. For this experiment, we use \(K=10\), \(u\in[0.01,10]\), and a 3-hour MILP time limit. For all considered map sizes, both methods obtain corridors with the minimum geometric length. The supported traffic densities obtained by the two methods remain close, whereas their computation times exhibit markedly different growth trends. The computational cost of the full MILP increases rapidly because the corridor and coverage variables are optimized over the entire grid. In contrast, the proposed method performs the corridor search using A* and restricts the subsequent integer optimization to the generated corridors, resulting in a much slower increase in computation time. For the larger maps, the MILP results are the best feasible solutions obtained within this limit. These results show that the proposed method maintains competitive solution quality while substantially reducing the computational burden as the problem size increases.

\begin{figure}[!htb]
    \centering
    \includegraphics[width=\columnwidth]{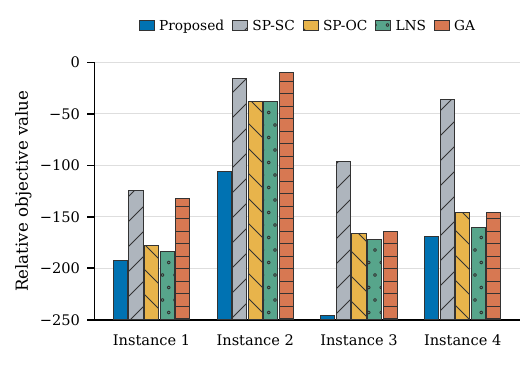}
    \caption{Relative objective values across four instances. }
    \label{fig:relative_objective}
\end{figure}

Fig.~\ref{fig:relative_objective} compares the relative objective
values achieved by the proposed method and the four benchmark schemes
over Instances 1--4. For each instance, the geometric shortest path
length is subtracted from the original objective of all methods as a
common offset, which preserves their performance ordering, and a
lower relative objective indicates a better solution. The proposed
method achieves the lowest relative objective in all four instances,
demonstrating a more favorable balance between corridor length and
supportable traffic density. SP--OC consistently outperforms SP--SC, demonstrating the benefit of optimizing BS coverage over the same candidate corridors. The additional gain achieved by the proposed method highlights the importance of accounting for BS resource sharing during
corridor search, as corridor selection affects how UAV traffic
can be distributed among BSs.

\begin{figure}[!htb]
    \centering
    \includegraphics[width=\columnwidth]{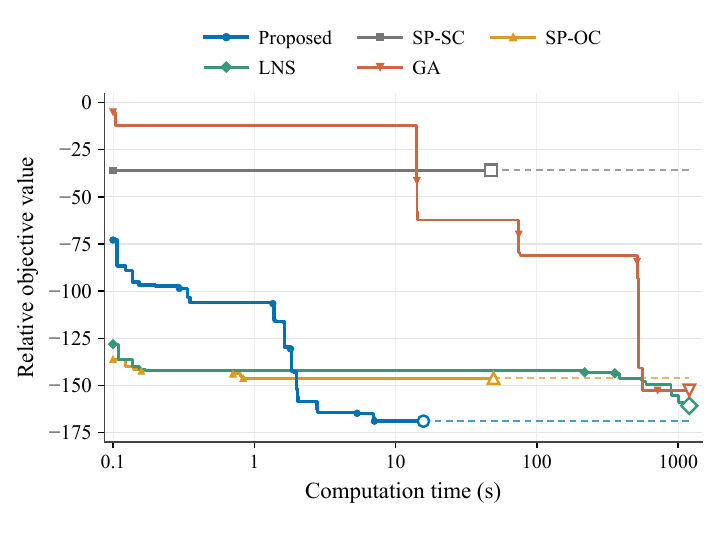}
    \caption{Relative objective versus computation time. Open markers indicate actual termination; lighter dashed
    lines retain the final solution after termination.}
    \label{fig:quality_time}
\end{figure}

Fig.~\ref{fig:quality_time} shows the evolution of the best feasible relative objective with computation time on Instance~4. The proposed method improves rapidly at the beginning of the search and reaches its final solution within a short computation time. SP--SC and SP--OC also terminate relatively early, but with higher objective values, whereas LNS and GA continue searching for much longer and improve more gradually. These results show that the proposed search strategy can identify high-quality corridor candidates efficiently without requiring a long search process.

\begin{figure}[!ht]
    \centering
    \includegraphics[width=\columnwidth]{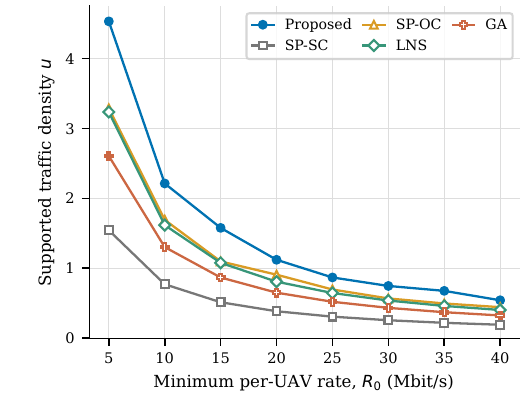}
    \caption{Supported aerial traffic density versus the minimum
    per-UAV rate requirement for Instance~0.}
    \label{fig:futian_rate_density}
\end{figure}

Fig.~\ref{fig:futian_rate_density} plots the supportable aerial traffic density versus the minimum per-UAV rate requirement \(R_0\). As \(R_0\) increases, the supported traffic density decreases for all methods because each UAV requires a larger share of the serving BS resources. The proposed method consistently achieves the highest traffic density over the entire range of \(R_0\), showing that the benefit of jointly optimizing the corridor and BS coverage persists under different rate requirements. The corridor and BS coverage are reoptimized for each value of \(R_0\), allowing the design to adapt to the changing communication constraints.

\section{Conclusion}
\label{sec:conclusion}

In this paper, we investigated the joint planning of air corridors
and BS coverage using CKMs to support dense UAV traffic in
terrestrial cellular networks. The proposed design jointly
determines the corridor layout and the corridor cells served by
each BS to balance corridor length and supportable UAV density
while accounting for BS resource sharing. Through an equivalent
reformulation of the rate constraints and exact linearization of
the products between UAV density and BS coverage variables,
we obtained an MILP formulation that can be solved to global
optimality. For large maps, we developed a scalable suboptimal
algorithm that uses Lagrangian relaxation to incorporate BS
resource demands into A* corridor search and then jointly
optimizes BS coverage and UAV density along the generated
corridors. Numerical results showed that the proposed algorithm
achieves better objective values than the considered benchmarks
and substantially reduces computation time compared with solving
the full MILP for larger maps. The results further demonstrated
that corridors with the same geometric length can support
markedly different UAV densities, highlighting the importance
of jointly planning corridor geometry and BS coverage under
resource sharing.

\appendices

\section{Proof of Theorem~\ref{thm:connectivity}}
\label{app:connectivity}

Consider the undirected graph formed by the selected grid cells, where two
vertices are connected if the corresponding cells are four-neighbor adjacent.
Constraints \eqref{eq:p1_endpoints}--\eqref{eq:p1_degree} ensure that the
departure and destination cells have degree one, while every other selected
cell has degree two. By the handshaking lemma, each connected component of an
undirected graph contains an even number of odd-degree vertices. Since the
departure and destination cells are the only selected vertices with odd
degree, they must belong to the same connected component. As this component
contains exactly two degree-one vertices and all its remaining vertices have
degree two, it forms a simple path connecting the departure and destination
cells. Any other connected component, if present, contains only degree-two
vertices and therefore forms an isolated cycle.

Suppose that a globally optimal solution contains an isolated cycle
whose grid-cell set is denoted by $\mathcal V$. Construct another solution by setting
\begin{equation}
    A_{i,j}=0,
    \quad
    x_{k,i,j}=0,
    \quad
    \forall (i,j)\in\mathcal V,\; k\in\Kset,
\end{equation}
while keeping $u$ unchanged and updating each $u_k$ according to
\eqref{eq:p1_load}. Since $\mathcal V$ is disconnected from the
departure-to-destination path, removing it preserves all connectivity
constraints. Moreover, the resulting BS loads satisfy
$u_k'\le u_k$ for all $k\in\Kset$. Hence, the per-UAV rate of every retained
covered grid cell does not decrease, and all communication constraints
remain feasible.

The removal of $\mathcal V$ leaves $u$ unchanged and reduces the objective
in \eqref{eq:p1_obj} by $\alpha_1|\mathcal V|>0,$ since $\alpha_1>0$. This contradicts the global optimality of the original
solution. Therefore, no isolated cycle can appear in a globally optimal
solution, and the selected grid cells form exactly one simple path connecting
the prescribed departure and destination cells.

\section{Proof of Theorem~\ref{thm:dual_convergence}}
\label{app:dual_convergence}

Fix a trial density $u^{(v)}$ for which Problem~(P3-$v$) is
feasible. For notational simplicity, write $\boldsymbol e^{(t)}$
in place of $\boldsymbol e^{(v,t)}$, and let
$\boldsymbol\lambda_v^\star$ denote an optimal solution of the
corresponding Lagrangian dual problem.

Since each Lagrangian subproblem is solved exactly, its resource
residual vector $\boldsymbol e^{(t)}$ is a supergradient of the
concave dual function $\phi_v$ at
$\boldsymbol\lambda^{(t)}$. Define the effective step size as
\begin{equation}
    a_t
    \triangleq
    \frac{\beta_0}
    {\sqrt{t+1}\max\{1,\|\boldsymbol e^{(t)}\|_2\}}.
\end{equation}
The multiplier update in \eqref{eq:multiplier_update} is therefore
projected supergradient ascent with step size $a_t$.

By the supergradient inequality and the nonexpansiveness of
projection onto the nonnegative orthant,
\begin{equation}
\begin{aligned}
&\|\boldsymbol\lambda^{(t+1)}
  -\boldsymbol\lambda_v^\star\|_2^2\\
&\quad\le
\|\boldsymbol\lambda^{(t)}
  -\boldsymbol\lambda_v^\star\|_2^2
-2a_t\bigl(
    \phi_v^\star
    -\phi_v(\boldsymbol\lambda^{(t)})
\bigr)
+a_t^2\|\boldsymbol e^{(t)}\|_2^2.
\end{aligned}
\end{equation}
Summing over $t=0,\ldots,T-1$ and using
$D_v^{(T)}\ge\phi_v(\boldsymbol\lambda^{(t)})$ for every $t<T$
gives the standard bound~\cite{bertsekas2015convex}
\begin{equation}
\phi_v^\star-D_v^{(T)}
\le
\frac{
\|\boldsymbol\lambda^{(0)}
-\boldsymbol\lambda_v^\star\|_2^2
+
\sum_{t=0}^{T-1}a_t^2\|\boldsymbol e^{(t)}\|_2^2
}{
2\sum_{t=0}^{T-1}a_t
}.
\label{eq:standard_subgradient_bound}
\end{equation}

Since the cell costs in \eqref{eq:lagrangian_cell_cost}
are positive, an optimal Lagrangian path contains no cycles
and hence at most $N^2$ cells. Combining this observation
with \eqref{eq:resource_residual} and
\eqref{eq:feasible_bs_set} gives
\begin{equation}
    -1\le e_k^{(t)}\le N^2-1,
    \qquad \forall k\in\mathcal K.
\end{equation}
Consequently,
\begin{equation}
    \|\boldsymbol e^{(t)}\|_2
    \le \psi
    \triangleq \sqrt K(N^2-1).
\end{equation}
Since $\psi\ge1$, the definition of $a_t$ yields
\begin{equation}
    a_t\ge
    \frac{\beta_0}{\psi\sqrt{t+1}}.
    \label{eq:PF2eq1}
\end{equation}
The normalization in $a_t$ gives
\begin{equation}
\begin{aligned}
a_t^2\|\boldsymbol e^{(t)}\|_2^2
&=
\frac{\beta_0^2}{t+1}
\frac{\|\boldsymbol e^{(t)}\|_2^2}
{\max\{1,\|\boldsymbol e^{(t)}\|_2\}^2}\\
&\le
\frac{\beta_0^2}{t+1}.
\end{aligned}
\label{eq:PF2eq2}
\end{equation}
Substituting \eqref{eq:PF2eq1} and \eqref{eq:PF2eq2} into
\eqref{eq:standard_subgradient_bound} yields
\begin{equation}
\phi_v^\star-D_v^{(T)}
\le
\frac{
\psi\left(
\|\boldsymbol\lambda^{(0)}
-\boldsymbol\lambda_v^\star\|_2^2
+
\beta_0^2\sum_{t=0}^{T-1}(t+1)^{-1}
\right)
}{
2\beta_0\sum_{t=0}^{T-1}(t+1)^{-1/2}
}.
\end{equation}

For a fixed problem instance and $\beta_0$, both $\psi$ and
$\|\boldsymbol\lambda^{(0)}
-\boldsymbol\lambda_v^\star\|_2^2$ are constants independent
of $T$. Using $\sum_{t=0}^{T-1}(t+1)^{-1}=O(\log(T+1))$
and $\sum_{t=0}^{T-1}(t+1)^{-1/2}=\Theta(\sqrt T)$, we obtain
\begin{equation}
    0\le
    \phi_v^\star-D_v^{(T)}
    =
    O\!\left(
        \frac{\log(T+1)}{\sqrt T}
    \right)
    \longrightarrow0.
\end{equation}
Therefore, $D_v^{(T)}\to\phi_v^\star$, proving both
\eqref{eq:dual_convergence} and
\eqref{eq:dual_convergence_rate}.
\hfill$\square$

\bibliographystyle{IEEEtran}
\bibliography{ref}

\end{document}